\documentclass[%
 reprint,
 amsmath,amssymb,
 aps,
pra,
]{revtex4-1}

\usepackage{graphicx}
\usepackage{dcolumn}
\usepackage{bm}
\usepackage{hyperref}

\begin{document}

\preprint{xxxxxxxxxxxxxx}

\title{Site-resolved imaging of a bosonic Mott insulator using ytterbium atoms}

\author{Martin Miranda}
\email{miranda.m.aa@m.titech.ac.jp}
\author{Ryotaro Inoue}
\author{Naoki Tambo}
\author{Mikio Kozuma}
\affiliation{Department of Physics, Tokyo Institute of Technology, 2-12-1 O-okayama, Meguro-ku, Tokyo 152-8550, Japan}

\date{\today}

\begin{abstract}
We demonstrate site-resolved imaging of a strongly correlated quantum system without relying on laser cooling techniques during fluorescence imaging. We observe the formation of Mott shells in the insulating regime and realize thermometry in an atomic cloud. This work proves the feasibility of the noncooled approach and opens the door to extending the detection technology to new atomic species.
\end{abstract}


\maketitle

\section{Introduction}

Since the creation of site-resolved fluorescence-imaging devices capable of observing a quantum gas trapped in a two-dimensional optical lattice\cite{greiner1}, there has been tremendous progress in the study of strongly correlated quantum systems. By observing the superfluid-to-Mott-insulator transition at the single atom level by using bosonic Rb atoms\cite{greiner2, bloch1}, scientists have been able to observe the phase transitions of interacting quantum Ising spins\cite{qgm_ising}, the dynamics of interacting quantum walkers\cite{qgm_randomwalk}, and magnon bound states\cite{qgm_magnon}. Moreover, the measurement of entanglement entropy has also been realized\cite{qgm_entropy}. Recently, detection technology has been expanded to fermionic Li\cite{li1, li2} and K\cite{k1,k2,k3}, culminating in the observation of a fermionic Mott insulator\cite{fermi_insulator1, fermi_insulator2} and long-range antiferromagnetic ordering\cite{antiferromagnetic}. These experiments are significant toward the understanding of d-wave superconductivity.
Improving the site-resolved imaging technology and extending it to new atomic species is an important step toward exploring a broader variety of strongly correlated phenomena.  Among the candidates for extending this technology, highly dipolar atoms such as Dy and Er are promising for studying the extended Bose-Hubbard model and its underlying exotic phases of matter\cite{extended_bhm1,extended_bhm2}.

The most challenging task in the realization of site-resolved fluorescence imaging is fulfilling the requirement that atoms stay localized within a site while their fluorescence is collected. The conventional method for achieving this is to perform laser cooling simultaneously with imaging. Different cooling methods have been applied in the past experiments, including polarization gradient cooling in the case of Rb\cite{greiner1}, Raman cooling for Li\cite{li1,li2} and K\cite{k1}, EIT cooling for K\cite{k2,k3}, and narrow-line optical molasses for Yb\cite{ytterbium_narrow}. Although these cooling techniques have proven to be effective for imaging with a fidelity near unity, the experimental setups are complicated and often   applicable only to a particular species. 

A promising alternative to laser cooling-based systems is to use a sufficiently deep optical potential and short exposure time. This method was demonstrated using Yb atoms in \cite{mm2}, where an optical lattice nearly resonant with a transition from the excited state was used to create a large light shift. The required deep potential was created by coupling the ground and excited states with an excitation beam. The main advantage of this experimental setup is that it only requires a single excitation beam and an available transition from the excited state and thus is readily extensible to new atomic species. 

Prior to this research, access to strongly correlated quantum systems was limited to the systems based on laser cooling, and it was not clear whether the noncooled approach would provide sufficient fidelity to access the required physics. Here we report the first direct observation of a Hubbard system using a noncooled site-resolved imaging device. We observe the shell structure of a bosonic ${}^{174}\textrm{Yb}$ Mott insulator with near unity fidelity, thus proving the effectiveness of this approach. We also go one step beyond the analysis in \cite{mm2} by considering loss rates that are not constant and by estimating an upper limit for the hopping probability.

This paper is organized as follows. Section \ref{section:experiment} briefly presents the experimental  and imaging setup. Section \ref{section:fidelity} focuses upon the estimation of loss and hopping probability. In Sec.~\ref{section:thermometry} we fit the reconstructed atomic-density distributions to measure the temperatures of atoms in the insulator regime. Finally, in Sec.~\ref{section:conclusion} we summarize and conclude our analysis.

\section{Experiment}
\label{section:experiment}

\begin{figure}
\includegraphics{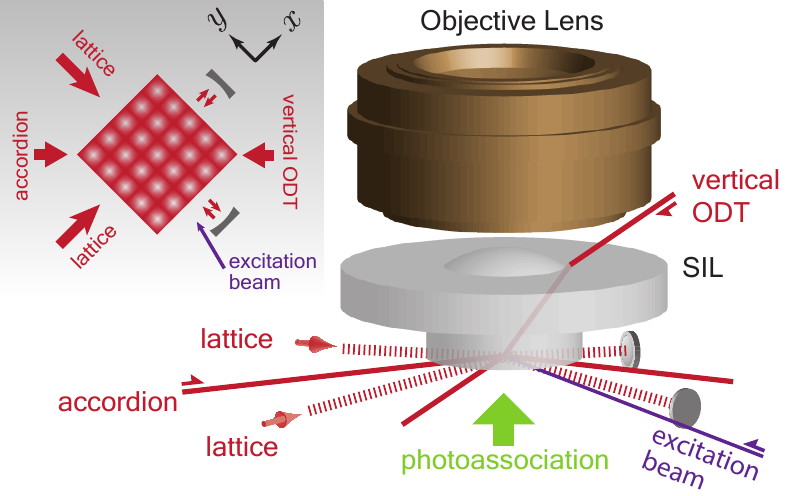}
\caption{\label{fig:figure1} Experimental setup. After creating a Bose-Einstein condensate and compressing it by using a combination of an optical accordion and a vertical ODT, a pair of retro-reflected lattice beams is introduced to load the atoms into the two-dimensional optical lattice. The system is then driven into a Mott insulating state by gradually ramping up the intensity of the lattice beams. A photo-association light is used to force inelastic light-assisted collisions on multiple occupied sites. Site-resolved imaging of the insulator is realized by irradiating a single excitation beam onto the atoms. No cooling mechanism is employed while the imaging system collects photons. }
\end{figure}

\begin{figure*}
\includegraphics{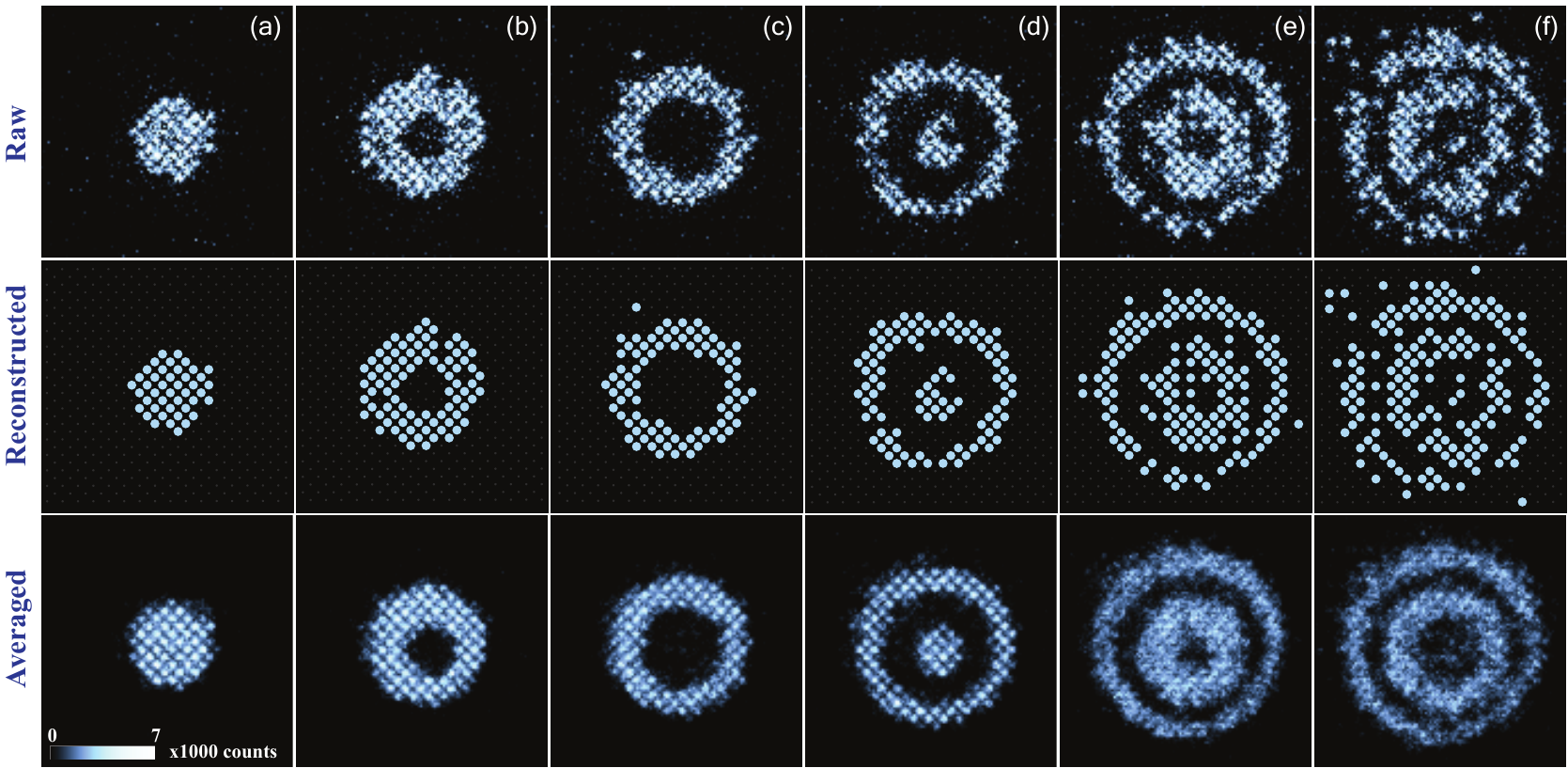}
\caption{\label{fig:figure_shells} Site-resolved imaging of a bosonic ${}^{174}\textrm{Yb}$ Mott insulator. Top row corresponds to raw images obtained with the emCCD camera during an exposure time of 40$\,\mu$s. The middle row shows the reconstructed atomic-density distribution obtained from the deconvolution algorithm. The estimated number of atoms in each image is (a) $45 \pm 7$, (b) $115 \pm 5$, (c) $179 \pm 13$, (d) $273 \pm 12$, (e) $486 \pm 21$ and (f) $592 \pm 27$. Bottom row shows the result of averaging 10 fluorescence raw images. Due to the presence of an external harmonic confinement in the optical lattices, the number of visible concentric Mott shells increases with the number of atoms. }
\end{figure*}

\subsection{Preparation of the Mott insulator}

We start our experiment by preparing a two-dimensional condensate of $5 \times 10^4$ bosonic ${}^{174}$Yb that is positioned $2.6\,\mu$m below the surface of a solid immersion lens (SIL). The SIL enables us to increase the resolution of the imaging system and additionally fix the position of the atoms relative to its flat surface. The procedure to create and compress the condensate utilizes the ``optical accordion'' technique described in \cite{mm1, mm2}. This technique comprises reflecting a laser beam from the flat substrate at a shallow angle to create a standing wave with tunable periodicity. In contrast to the procedure explained in \cite{mm2}, which utilizes a combination of two orthogonal optical accordions and one vertical optical dipole trap (ODT), here we opt to perform evaporative cooling using only one accordion beam and the vertical ODT (see Fig. 1). Both beams have a wavelength of $1080\,$nm and propagate in the $x+y=0$ vertical plane. After compressing the condensate, we perform a second evaporative cooling by reducing the accordion beam power over $9\,$s. We control the number of atoms loaded into the two-dimensional optical lattice by adjusting the final power of the accordion beam.


To load the atoms into the two-dimensional optical lattice, we use an additional pair of beams (wavelength 1080 nm) propagating in the orthogonal $x=0$ and $y=0$ planes. The lattice beams are reflected from the SIL at the same angle as the optical accordion and retro-reflected by using a concave mirror with a $50\,\textrm{mm}$ radius of curvature. This creates a two-dimensional lattice with spacing $a_\textrm{lat}=543.5\,\textrm{nm}$ in the $x$-$y$ plane and a standing wave with $4.8\mu\textrm{m} $ spacing in the $z$ direction. The lattice beams have an elliptical cross-section, with waists of $26\, \mu\textrm{m}$ and $52\, \mu\textrm{m}$ in the $z$ and $x(y)$ directions, respectively. We load the atoms into the lattice by ramping up the potential depth to $6.5\, E_r$ over $2\, \textrm{s}$ while decreasing the intensity of the vertical and accordion beams. At this point, the atoms are in the superfluid regime, which we confirm by the presence of sharp interference peaks in the momentum distribution\cite{Greiner2002}. We further increase the lattice depth to $26 \, E_r$ over $1\,\textrm{s}$ by using a smooth S-shaped curve to induce a phase transition into a Mott insulator\footnote{The superfluid-to-Mott-insulator transition is expected to occur at a lattice depth of $14\, E_r$. At the transition point, the tunneling rate is $J/h=7.7\,\text{Hz}$, the interaction energy is $U/k_B=6.1\,\text{nK}$ and the lattice transverse confinement is $\omega / 2\pi = 35\,\text{Hz}$. At  $26\, E_r$ the tunneling rate becomes negligible compared with the on-site interaction ($U/J \approx 240$) and fluctuations in the atom number are drastically reduced.}.

\subsection{Site-resolved fluorescence imaging}


We employ the photo-association (PA) technique to remove pairs of atoms in multiply occupied sites and realize parity measurement of the number density\cite{556_PA,556_PA2}. The PA laser is red-detuned by $301 \, \textrm{MHz}$ from the ${}^{1}\textrm{S}_{0}-{}^{3}\textrm{P}_{1}$ atomic transition at $556\,\textrm{nm}$. Pairs of atoms decay over $34(22)\,\mu\textrm{s}$ when an optical lattice depth of $1200\, E_r$ and a PA laser beam intensity of $0.75\, \textrm{W}/\textrm{cm}^2$ is used. For this experiment, we ramp up the lattice depth in $10\, \textrm{ms}$ and irradiate the PA beam for a period of $2\,\textrm{ms}$. This is expected to eliminate $>99\%$ of the atomic pairs while only producing an average of $\sim 0.5$ photon scatterings in the rest of the atoms. 

Finally, we obtain site-resolved imaging of a Mott insulator by further increasing the lattice depth to $3200\, E_r$ over $5\,\textrm{ms}$ and irradiating an excitation beam (wavelength $399\,\textrm{nm}$, intensity $65\,\textrm{W}/\textrm{cm}^2$) upon the atoms for $40\, \mu\textrm{s}$. The scattered photons are collected by a high-resolution optical system (numerical aperture 0.81, magnification 110X) composed of the SIL and an objective lens and then focused into an emCCD camera (Andor iXon Ultra 888). The top row of Fig. \ref{fig:figure_shells} shows the obtained raw images for an increasing number of atoms in the trap. The observed concentric shells correspond to a fixed number of atoms in each shell, which is the characteristic structure of a Mott insulator under harmonic confinement\cite{shell1,shell2}. To reconstruct the atomic-density distribution $n_\text{det}(\mathbf{r})$, we first estimate the total fluorescence at each site by employing a computer algorithm based on deconvolution. The obtained total fluorescence is then compared with a previously determined fluorescence threshold to determine which site was occupied. The middle row in Fig.~\ref{fig:figure_shells} shows the estimated density distribution corresponding to each of the images on top. Images in the bottom row are the result of averaging 10 fluorescence raw images.



\section{Fidelity}
\label{section:fidelity}

Here, we study the fidelity of the imaging system. In site-resolved imaging devices relying on cooling-based schemes to pin the atoms during the imaging process, the conventional method for estimating hopping and loss effects is to take two successive fluorescence images and compare their observed atomic-density distributions. As atoms are thermally in equilibrium during laser cooling, hopping and loss rates are constant. Thus, the comparison method provides a good estimation of both rates. In the case of the noncooled approach, the temperature of atoms during imaging is not constant but continuously increasing. This results in a number of trapped atoms that decays non-exponentially.

\begin{figure}
\includegraphics{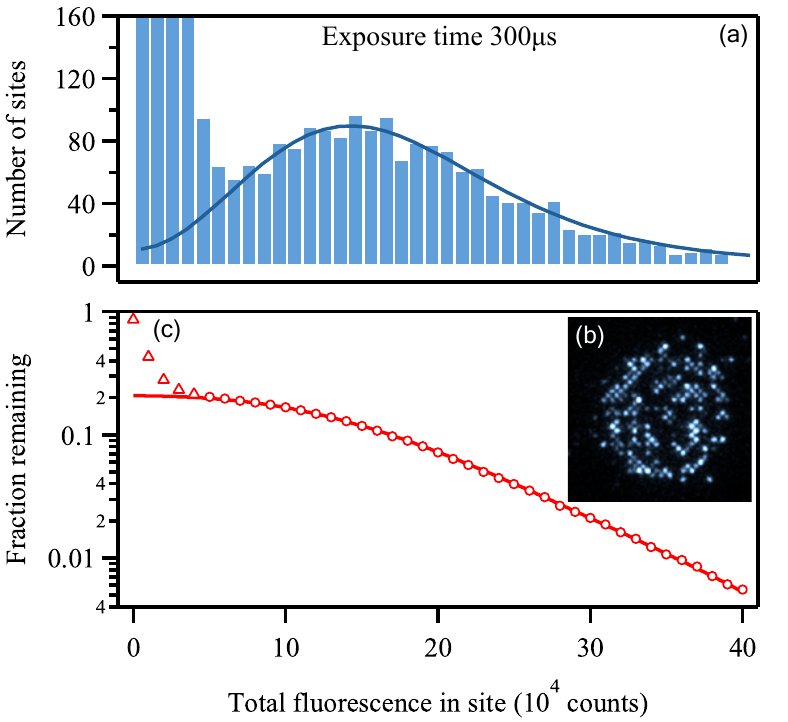}
\caption{\label{fig:figure_loss} Estimation of loss effects. (a) Histogram for the number of sites as a function of the total fluorescence at each site using 10 images taken at an exposure of 300 $\,\mu\textrm{s}$. The left peak corresponds to empty sites that are affected by stray background light. (b) Typical raw image. (c) Computed complementary cumulative function of the histogram, shown as a semi-log plot. Circular points are fitted with a simulation (solid line) to obtain the percentage of occupied sites. Triangular points are not considered in the fitting because they include the contribution of empty sites. }
\end{figure}

\subsection{Loss effects}

To estimate the loss effects, we obtain fluorescence images by employing the same procedure as in Fig. \ref{fig:figure_shells} but using a longer exposure time ($300\,\mu\textrm{s}$ instead of $40\,\mu\textrm{s}$). A typical observed image is shown in Fig. \ref{fig:figure_loss}(b). Figure \ref{fig:figure_loss}(a) shows the computed histogram for the number of sites as a function of the total fluorescence at each site. Note that the left peak of the histogram is determined by background noise on empty sites, which is caused by stray light from the excitation beam. From the histogram we compute the complementary cumulative distribution as shown in Fig.~\ref{fig:figure_loss}(c).  For a large number of fluorescence counts, the background noise becomes negligible and the distribution is determined only by the fluorescence on occupied sites (circular points in Fig.~\ref{fig:figure_loss}(c)). In accordance with the law of large numbers, the distribution of occupied sites is expected to be equivalent to the probability, $P_S(N_\text{F})$, of an atom surviving after emitting $N_\text{F}$ fluorescence counts, that is multiplied by the percentage of occupied sites. In the case of a cooling-based scheme, $P_S(N_\text{F})$ will decay exponentially as atoms have a constant temperature during imaging, resulting in a straight line on the semi-log plot. In contrast, for the noncooled approach presented here, we observe a non-exponential decay. We fit the distribution of occupied sites with a known $P_S$ obtained by simulation. This simulation considers losses due to heating and also light-induced excitations from the optical lattice\cite{mm2}. From the fitting, we estimate that $21\%$ of the sites are initially occupied. The solid line in Fig.~\ref{fig:figure_loss}(c) shows the fitting result. We find a remarkable agreement between the experimental data and the simulation, even at very large fluorescence counts. As a reference, we have also included the estimated curve for the histogram (solid line in Fig.~\ref{fig:figure_loss}(a)), which can be computed directly from $P_S$. We can then calculate the loss rate and percentage of lost atoms from the derivative and complement of $P_S$, respectively.

\begin{figure}
\includegraphics{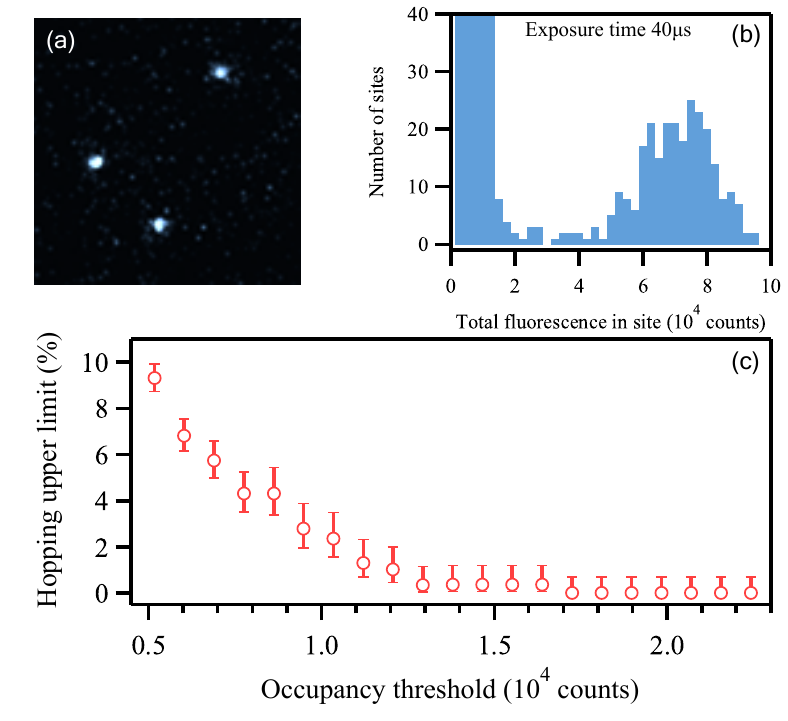}
\caption{\label{fig:figure_hopping} Estimation of hopping effects. (a) Typical raw image with a sparse population taken at an exposure time of $40\,\mu\textrm{s}$. (b) Histogram of the total fluorescence per site using $100$ images containing an average of $2.6$ atoms per image. Left and right peaks correspond to empty and occupied sites, respectively. (c) Upper limit of the hopping probability as a function of the occupancy threshold. Error bars denote $68\%$ Clopper-Pearson confidence intervals. }
\end{figure}

\subsection{Hopping effects}

Hopping effects are estimated using 100 images (exposure time $40\,\mu\textrm{s}$) of lattices with sparse and low populations, i.e., $2.6$ atoms on an average (see Fig.~\ref{fig:figure_hopping}(a)). These samples are prepared by ramping down the lattice depth to $4\, E_r$ over $1\,\textrm{ms}$ followed by a $0.5\,\textrm{s}$ holding time. The shallow lattice depth allows atoms to disperse randomly along the lattice while the total number of atoms is reduced. The lattice depth is then ramped up to $26 E_r$ over $10\,\textrm{ms}$ to pin the positions of the atoms in the lattice and later imaged in the same way as the Mott insulator.

For each reconstructed image, we determine which sites are occupied or not by comparing the total fluorescence in each site with an occupancy threshold. We then calculate the number of total occupied sites and the number of groups of two adjacent occupied sites in each image. Adjacent occupied sites are the result of either hopping events, a pair of atoms randomly occupying two adjacent sites\footnote{We estimate that the probability of finding two atoms occupying adjacent sites is $0.5\%$ for a square lattice comprised of $30\times30$ sites.}, or background events. For different choices of the occupancy threshold, the probability of finding two adjacent sites that are occupied then establishes an upper limit for the hopping probability, as shown in Fig.~\ref{fig:figure_hopping}(c).

\subsection{Loss and hopping probability}

We set the occupancy threshold to $2 \times 10^4$ fluorescence counts. For this threshold, the loss and hopping probabilities are $1.8\%$ and less than $0.7\%$, respectively. The low hopping probability is a characteristic of the non-cooled imaging system, because atoms that become heated are rapidly accelerated by the radiative force exerted by the excitation beam and very rarely emit a sufficient number of photons in the neighboring sites for these sites to be considered as occupied. 


\section{Thermometry}
\label{section:thermometry}


\begin{figure}
\includegraphics{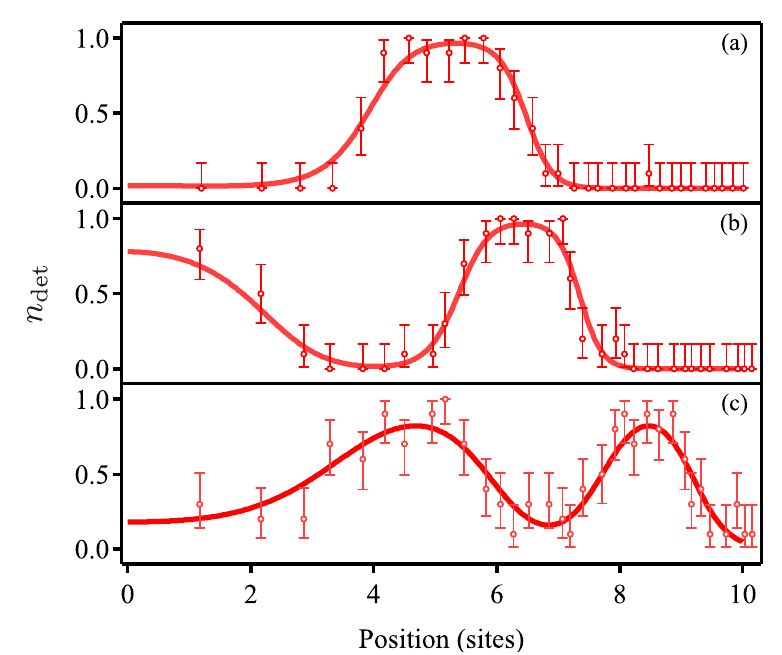}
\caption{\label{fig:figure3} Temperature measurement of a Mott insulator. The reconstructed atomic-density distribution is averaged azimuthally to obtain the radial profiles (points) and fitted using the grand-canonical ensemble described by the Bose-Hubbard model under the zero-tunneling approximation (solid lines). Error bars denotes $68\%$ Clopper-Pearson confidence intervals. The experimental data for (a)-(c) correspond to the reconstructed density distributions in Fig.~\ref{fig:figure_shells}(c), (d) and (f), respectively. From each fitting, the global chemical potential $\mu_0$, temperature $T$, and Mott radius $r_0$ are extracted. }
\end{figure}

Finally, we measure the temperature of the atomic cloud by analyzing the reconstructed density distribution. The Bose-Hubbard model describes the behavior of atoms trapped in a two-dimensional optical lattice with harmonic confinement. When the tunneling rate, $J$, is sufficiently smaller than the on-site interaction energy $U$ ($U/J \gg 16.7$)\cite{bhm1}, the number density after parity projection, $n_\textrm{det}(\mathbf{r})$, is approximately (zero-tunnelling approximation)\cite{bloch1}:
\begin{equation*}
n_\textrm{det}(\mathbf{r}) = \frac{1}{Z} \sum_{n=1}^\infty \text{mod}_2(n) \exp \left[ \frac{\mu(r) n - E_n}{k_B T} \right],
\end{equation*}
where $Z$ is the grand canonical partition function, $\mu$ is the local chemical potential, $T$ is the temperature, $k_B$ is the Boltzmann constant, and $E_n = U n(n-1)/2$ is the interaction energy for a site occupied by $n$ atoms. We apply the local density approximation $\mu = \mu_0 - 0.5 m \omega^2 r^2$ where $\mu_0$ is the global chemical potential and $\omega$ is the trap frequency of the harmonic confinement. Note that we consider an azimuthally symmetric function because we measured negligible ellipticity in our trap geometry. 

We average the reconstructed number density azimuthally and then fit the result with the theoretical $n_\textrm{det}(\mathbf{r})$  (see Fig.~\ref{fig:figure3}), taking the loss effects into account. From the fitting, we extract the parameters $\mu_0 / U$ and  $T / U$, as well as the Mott shell radius $r_0=\sqrt{ 2U / m\omega^2 }$. This yield the parameters $T =0.10(1) U/k_B$, $\mu_0 = 1.56(6) U$, $r_0 = 5.1(1) a_\textrm{lat}$ for (a), $T =0.10(1) U/k_B$, $\mu_0 = 2.14(4) U$, $r_0 = 5.0(1) a_\textrm{lat}$ for (b) and $T =0.21(2) U/k_B$, $\mu_0 = 3.38(7) U$, $r_0 = 5.0(1) a_\textrm{lat}$ for (c). From the extracted parameters we also calculate the entropy per atom resulting in $0.32(6) k_B$ for (a),  $0.28(4) k_B$ for (b) and $0.34(3) k_B$ for (c).

The errors in the computed parameters are caused by the limited number of sites used in the azimuthal averages, which is reflected by the size of the error bars on the experimental data. Hopping effects are very small and produce a negligible error in the measurement. Increasing the accuracy of this thermometer would require the use of traps having smaller $\omega$ or increasing $U$ by using larger s-wave scattering lengths.



\section{Conclusions}
\label{section:conclusion}

In conclusion, we have demonstrated the first site-resolved observation of a Mott insulator by using a noncooled method. 
This approach is robust against mechanical instabilities in the optical system owing to the short exposure time used during imaging. The simplicity of the setup that uses only one excitation beam that does not require retro-reflection makes it readily applicable to other species. In particular, lanthanoid atoms benefit from the noncooled method as they have a large mass, which results in small recoil energies ensuring small losses. We have also presented a method for estimating the loss and hopping effects and found that our system has comparable fidelity to that its laser cooling-based counterparts. Our results are promising for the study of the Fermi-Hubbard model using a Yb gas in its generalized SU(N) form\cite{fermi_sun1,fermi_sun2}.

\begin{acknowledgments}
This work was supported by JST CREST (Grant Number JPMJCR16N4), JSPS KAKENHI (Grant Numbers JP17H02934, JP26800212, JP16F16029, and JP16K05498), the Tokyo Tech Suematsu Award, and the Research
Foundation for Opto-Science and Technology. One of the authors (M.M.) is supported in part by the Japan Society for the Promotion of Science.
\end{acknowledgments}


\bibliography{third} 

\providecommand{\noopsort}[1]{}\providecommand{\singleletter}[1]{#1}%
\begin{thebibliography}{30}%
\makeatletter
\providecommand \@ifxundefined [1]{%
 \@ifx{#1\undefined}
}%
\providecommand \@ifnum [1]{%
 \ifnum #1\expandafter \@firstoftwo
 \else \expandafter \@secondoftwo
 \fi
}%
\providecommand \@ifx [1]{%
 \ifx #1\expandafter \@firstoftwo
 \else \expandafter \@secondoftwo
 \fi
}%
\providecommand \natexlab [1]{#1}%
\providecommand \enquote  [1]{``#1''}%
\providecommand \bibnamefont  [1]{#1}%
\providecommand \bibfnamefont [1]{#1}%
\providecommand \citenamefont [1]{#1}%
\providecommand \href@noop [0]{\@secondoftwo}%
\providecommand \href [0]{\begingroup \@sanitize@url \@href}%
\providecommand \@href[1]{\@@startlink{#1}\@@href}%
\providecommand \@@href[1]{\endgroup#1\@@endlink}%
\providecommand \@sanitize@url [0]{\catcode `\\12\catcode `\$12\catcode
  `\&12\catcode `\#12\catcode `\^12\catcode `\_12\catcode `\%12\relax}%
\providecommand \@@startlink[1]{}%
\providecommand \@@endlink[0]{}%
\providecommand \url  [0]{\begingroup\@sanitize@url \@url }%
\providecommand \@url [1]{\endgroup\@href {#1}{\urlprefix }}%
\providecommand \urlprefix  [0]{URL }%
\providecommand \Eprint [0]{\href }%
\providecommand \doibase [0]{http://dx.doi.org/}%
\providecommand \selectlanguage [0]{\@gobble}%
\providecommand \bibinfo  [0]{\@secondoftwo}%
\providecommand \bibfield  [0]{\@secondoftwo}%
\providecommand \translation [1]{[#1]}%
\providecommand \BibitemOpen [0]{}%
\providecommand \bibitemStop [0]{}%
\providecommand \bibitemNoStop [0]{.\EOS\space}%
\providecommand \EOS [0]{\spacefactor3000\relax}%
\providecommand \BibitemShut  [1]{\csname bibitem#1\endcsname}%
\let\auto@bib@innerbib\@empty
\bibitem [{\citenamefont {Bakr}\ \emph {et~al.}(2009)\citenamefont {Bakr},
  \citenamefont {Gillen}, \citenamefont {Peng}, \citenamefont {Folling},\ and\
  \citenamefont {Greiner}}]{greiner1}%
  \BibitemOpen
  \bibfield  {author} {\bibinfo {author} {\bibfnamefont {W.~S.}\ \bibnamefont
  {Bakr}}, \bibinfo {author} {\bibfnamefont {J.~I.}\ \bibnamefont {Gillen}},
  \bibinfo {author} {\bibfnamefont {A.}~\bibnamefont {Peng}}, \bibinfo {author}
  {\bibfnamefont {S.}~\bibnamefont {Folling}}, \ and\ \bibinfo {author}
  {\bibfnamefont {M.}~\bibnamefont {Greiner}},\ }\href {\doibase
  10.1038/nature08482} {\bibfield  {journal} {\bibinfo  {journal} {Nature}\
  }\textbf {\bibinfo {volume} {462}},\ \bibinfo {pages} {74} (\bibinfo {year}
  {2009})}\BibitemShut {NoStop}%
\bibitem [{\citenamefont {Bakr}\ \emph {et~al.}(2010)\citenamefont {Bakr},
  \citenamefont {Peng}, \citenamefont {Tai}, \citenamefont {Ma}, \citenamefont
  {Simon}, \citenamefont {Gillen}, \citenamefont {F{\"o}lling}, \citenamefont
  {Pollet},\ and\ \citenamefont {Greiner}}]{greiner2}%
  \BibitemOpen
  \bibfield  {author} {\bibinfo {author} {\bibfnamefont {W.~S.}\ \bibnamefont
  {Bakr}}, \bibinfo {author} {\bibfnamefont {A.}~\bibnamefont {Peng}}, \bibinfo
  {author} {\bibfnamefont {M.~E.}\ \bibnamefont {Tai}}, \bibinfo {author}
  {\bibfnamefont {R.}~\bibnamefont {Ma}}, \bibinfo {author} {\bibfnamefont
  {J.}~\bibnamefont {Simon}}, \bibinfo {author} {\bibfnamefont {J.~I.}\
  \bibnamefont {Gillen}}, \bibinfo {author} {\bibfnamefont {S.}~\bibnamefont
  {F{\"o}lling}}, \bibinfo {author} {\bibfnamefont {L.}~\bibnamefont {Pollet}},
  \ and\ \bibinfo {author} {\bibfnamefont {M.}~\bibnamefont {Greiner}},\ }\href
  {\doibase 10.1126/science.1192368} {\bibfield  {journal} {\bibinfo  {journal}
  {Science}\ }\textbf {\bibinfo {volume} {329}},\ \bibinfo {pages} {547}
  (\bibinfo {year} {2010})}\BibitemShut {NoStop}%
\bibitem [{\citenamefont {Sherson}\ \emph {et~al.}(2010)\citenamefont
  {Sherson}, \citenamefont {Weitenberg}, \citenamefont {Endres}, \citenamefont
  {Cheneau}, \citenamefont {Bloch},\ and\ \citenamefont {Kuhr}}]{bloch1}%
  \BibitemOpen
  \bibfield  {author} {\bibinfo {author} {\bibfnamefont {J.~F.}\ \bibnamefont
  {Sherson}}, \bibinfo {author} {\bibfnamefont {C.}~\bibnamefont {Weitenberg}},
  \bibinfo {author} {\bibfnamefont {M.}~\bibnamefont {Endres}}, \bibinfo
  {author} {\bibfnamefont {M.}~\bibnamefont {Cheneau}}, \bibinfo {author}
  {\bibfnamefont {I.}~\bibnamefont {Bloch}}, \ and\ \bibinfo {author}
  {\bibfnamefont {S.}~\bibnamefont {Kuhr}},\ }\href {\doibase
  10.1038/nature09378} {\bibfield  {journal} {\bibinfo  {journal} {Nature}\
  }\textbf {\bibinfo {volume} {467}},\ \bibinfo {pages} {68} (\bibinfo {year}
  {2010})}\BibitemShut {NoStop}%
\bibitem [{\citenamefont {Simon}\ \emph {et~al.}(2011)\citenamefont {Simon},
  \citenamefont {Bakr}, \citenamefont {Ma}, \citenamefont {Tai}, \citenamefont
  {Preiss},\ and\ \citenamefont {Greiner}}]{qgm_ising}%
  \BibitemOpen
  \bibfield  {author} {\bibinfo {author} {\bibfnamefont {J.}~\bibnamefont
  {Simon}}, \bibinfo {author} {\bibfnamefont {W.~S.}\ \bibnamefont {Bakr}},
  \bibinfo {author} {\bibfnamefont {R.}~\bibnamefont {Ma}}, \bibinfo {author}
  {\bibfnamefont {M.~E.}\ \bibnamefont {Tai}}, \bibinfo {author} {\bibfnamefont
  {P.~M.}\ \bibnamefont {Preiss}}, \ and\ \bibinfo {author} {\bibfnamefont
  {M.}~\bibnamefont {Greiner}},\ }\href {\doibase 10.1038/nature09994}
  {\bibfield  {journal} {\bibinfo  {journal} {Nature}\ }\textbf {\bibinfo
  {volume} {472}},\ \bibinfo {pages} {307} (\bibinfo {year}
  {2011})}\BibitemShut {NoStop}%
\bibitem [{\citenamefont {Preiss}\ \emph {et~al.}(2015)\citenamefont {Preiss},
  \citenamefont {Ma}, \citenamefont {Tai}, \citenamefont {Lukin}, \citenamefont
  {Rispoli}, \citenamefont {Zupancic}, \citenamefont {Lahini}, \citenamefont
  {Islam},\ and\ \citenamefont {Greiner}}]{qgm_randomwalk}%
  \BibitemOpen
  \bibfield  {author} {\bibinfo {author} {\bibfnamefont {P.~M.}\ \bibnamefont
  {Preiss}}, \bibinfo {author} {\bibfnamefont {R.}~\bibnamefont {Ma}}, \bibinfo
  {author} {\bibfnamefont {M.~E.}\ \bibnamefont {Tai}}, \bibinfo {author}
  {\bibfnamefont {A.}~\bibnamefont {Lukin}}, \bibinfo {author} {\bibfnamefont
  {M.}~\bibnamefont {Rispoli}}, \bibinfo {author} {\bibfnamefont
  {P.}~\bibnamefont {Zupancic}}, \bibinfo {author} {\bibfnamefont
  {Y.}~\bibnamefont {Lahini}}, \bibinfo {author} {\bibfnamefont
  {R.}~\bibnamefont {Islam}}, \ and\ \bibinfo {author} {\bibfnamefont
  {M.}~\bibnamefont {Greiner}},\ }\href {\doibase 10.1126/science.1260364}
  {\bibfield  {journal} {\bibinfo  {journal} {Science}\ }\textbf {\bibinfo
  {volume} {347}},\ \bibinfo {pages} {1229} (\bibinfo {year}
  {2015})}\BibitemShut {NoStop}%
\bibitem [{\citenamefont {Fukuhara}\ \emph {et~al.}(2013)\citenamefont
  {Fukuhara}, \citenamefont {Schausz}, \citenamefont {Endres}, \citenamefont
  {Hild}, \citenamefont {Cheneau}, \citenamefont {Bloch},\ and\ \citenamefont
  {Gross}}]{qgm_magnon}%
  \BibitemOpen
  \bibfield  {author} {\bibinfo {author} {\bibfnamefont {T.}~\bibnamefont
  {Fukuhara}}, \bibinfo {author} {\bibfnamefont {P.}~\bibnamefont {Schausz}},
  \bibinfo {author} {\bibfnamefont {M.}~\bibnamefont {Endres}}, \bibinfo
  {author} {\bibfnamefont {S.}~\bibnamefont {Hild}}, \bibinfo {author}
  {\bibfnamefont {M.}~\bibnamefont {Cheneau}}, \bibinfo {author} {\bibfnamefont
  {I.}~\bibnamefont {Bloch}}, \ and\ \bibinfo {author} {\bibfnamefont
  {C.}~\bibnamefont {Gross}},\ }\href {http://dx.doi.org/10.1038/nature12541}
  {\bibfield  {journal} {\bibinfo  {journal} {Nature}\ }\textbf {\bibinfo
  {volume} {502}},\ \bibinfo {pages} {76} (\bibinfo {year} {2013})},\ \bibinfo
  {note} {letter}\BibitemShut {NoStop}%
\bibitem [{\citenamefont {Islam}\ \emph {et~al.}(2015)\citenamefont {Islam},
  \citenamefont {Ma}, \citenamefont {Preiss}, \citenamefont {Eric~Tai},
  \citenamefont {Lukin}, \citenamefont {Rispoli},\ and\ \citenamefont
  {Greiner}}]{qgm_entropy}%
  \BibitemOpen
  \bibfield  {author} {\bibinfo {author} {\bibfnamefont {R.}~\bibnamefont
  {Islam}}, \bibinfo {author} {\bibfnamefont {R.}~\bibnamefont {Ma}}, \bibinfo
  {author} {\bibfnamefont {P.~M.}\ \bibnamefont {Preiss}}, \bibinfo {author}
  {\bibfnamefont {M.}~\bibnamefont {Eric~Tai}}, \bibinfo {author}
  {\bibfnamefont {A.}~\bibnamefont {Lukin}}, \bibinfo {author} {\bibfnamefont
  {M.}~\bibnamefont {Rispoli}}, \ and\ \bibinfo {author} {\bibfnamefont
  {M.}~\bibnamefont {Greiner}},\ }\href {http://dx.doi.org/10.1038/nature15750}
  {\bibfield  {journal} {\bibinfo  {journal} {Nature}\ }\textbf {\bibinfo
  {volume} {528}},\ \bibinfo {pages} {77} (\bibinfo {year} {2015})},\ \bibinfo
  {note} {article}\BibitemShut {NoStop}%
\bibitem [{\citenamefont {Parsons}\ \emph {et~al.}(2015)\citenamefont
  {Parsons}, \citenamefont {Huber}, \citenamefont {Mazurenko}, \citenamefont
  {Chiu}, \citenamefont {Setiawan}, \citenamefont {Wooley-Brown}, \citenamefont
  {Blatt},\ and\ \citenamefont {Greiner}}]{li1}%
  \BibitemOpen
  \bibfield  {author} {\bibinfo {author} {\bibfnamefont {M.~F.}\ \bibnamefont
  {Parsons}}, \bibinfo {author} {\bibfnamefont {F.}~\bibnamefont {Huber}},
  \bibinfo {author} {\bibfnamefont {A.}~\bibnamefont {Mazurenko}}, \bibinfo
  {author} {\bibfnamefont {C.~S.}\ \bibnamefont {Chiu}}, \bibinfo {author}
  {\bibfnamefont {W.}~\bibnamefont {Setiawan}}, \bibinfo {author}
  {\bibfnamefont {K.}~\bibnamefont {Wooley-Brown}}, \bibinfo {author}
  {\bibfnamefont {S.}~\bibnamefont {Blatt}}, \ and\ \bibinfo {author}
  {\bibfnamefont {M.}~\bibnamefont {Greiner}},\ }\href {\doibase
  10.1103/PhysRevLett.114.213002} {\bibfield  {journal} {\bibinfo  {journal}
  {Phys. Rev. Lett.}\ }\textbf {\bibinfo {volume} {114}},\ \bibinfo {pages}
  {213002} (\bibinfo {year} {2015})}\BibitemShut {NoStop}%
\bibitem [{\citenamefont {Omran}\ \emph {et~al.}(2015)\citenamefont {Omran},
  \citenamefont {Boll}, \citenamefont {Hilker}, \citenamefont {Kleinlein},
  \citenamefont {Salomon}, \citenamefont {Bloch},\ and\ \citenamefont
  {Gross}}]{li2}%
  \BibitemOpen
  \bibfield  {author} {\bibinfo {author} {\bibfnamefont {A.}~\bibnamefont
  {Omran}}, \bibinfo {author} {\bibfnamefont {M.}~\bibnamefont {Boll}},
  \bibinfo {author} {\bibfnamefont {T.~A.}\ \bibnamefont {Hilker}}, \bibinfo
  {author} {\bibfnamefont {K.}~\bibnamefont {Kleinlein}}, \bibinfo {author}
  {\bibfnamefont {G.}~\bibnamefont {Salomon}}, \bibinfo {author} {\bibfnamefont
  {I.}~\bibnamefont {Bloch}}, \ and\ \bibinfo {author} {\bibfnamefont
  {C.}~\bibnamefont {Gross}},\ }\href {\doibase 10.1103/PhysRevLett.115.263001}
  {\bibfield  {journal} {\bibinfo  {journal} {Phys. Rev. Lett.}\ }\textbf
  {\bibinfo {volume} {115}},\ \bibinfo {pages} {263001} (\bibinfo {year}
  {2015})}\BibitemShut {NoStop}%
\bibitem [{\citenamefont {Cheuk}\ \emph {et~al.}(2015)\citenamefont {Cheuk},
  \citenamefont {Nichols}, \citenamefont {Okan}, \citenamefont {Gersdorf},
  \citenamefont {Ramasesh}, \citenamefont {Bakr}, \citenamefont {Lompe},\ and\
  \citenamefont {Zwierlein}}]{k1}%
  \BibitemOpen
  \bibfield  {author} {\bibinfo {author} {\bibfnamefont {L.~W.}\ \bibnamefont
  {Cheuk}}, \bibinfo {author} {\bibfnamefont {M.~A.}\ \bibnamefont {Nichols}},
  \bibinfo {author} {\bibfnamefont {M.}~\bibnamefont {Okan}}, \bibinfo {author}
  {\bibfnamefont {T.}~\bibnamefont {Gersdorf}}, \bibinfo {author}
  {\bibfnamefont {V.~V.}\ \bibnamefont {Ramasesh}}, \bibinfo {author}
  {\bibfnamefont {W.~S.}\ \bibnamefont {Bakr}}, \bibinfo {author}
  {\bibfnamefont {T.}~\bibnamefont {Lompe}}, \ and\ \bibinfo {author}
  {\bibfnamefont {M.~W.}\ \bibnamefont {Zwierlein}},\ }\href {\doibase
  10.1103/PhysRevLett.114.193001} {\bibfield  {journal} {\bibinfo  {journal}
  {Phys. Rev. Lett.}\ }\textbf {\bibinfo {volume} {114}},\ \bibinfo {pages}
  {193001} (\bibinfo {year} {2015})}\BibitemShut {NoStop}%
\bibitem [{\citenamefont {Haller}\ \emph {et~al.}(2015)\citenamefont {Haller},
  \citenamefont {Hudson}, \citenamefont {Kelly}, \citenamefont {Cotta},
  \citenamefont {Peaudecerf}, \citenamefont {Bruce},\ and\ \citenamefont
  {Kuhr}}]{k2}%
  \BibitemOpen
  \bibfield  {author} {\bibinfo {author} {\bibfnamefont {E.}~\bibnamefont
  {Haller}}, \bibinfo {author} {\bibfnamefont {J.}~\bibnamefont {Hudson}},
  \bibinfo {author} {\bibfnamefont {A.}~\bibnamefont {Kelly}}, \bibinfo
  {author} {\bibfnamefont {D.~A.}\ \bibnamefont {Cotta}}, \bibinfo {author}
  {\bibfnamefont {B.}~\bibnamefont {Peaudecerf}}, \bibinfo {author}
  {\bibfnamefont {G.~D.}\ \bibnamefont {Bruce}}, \ and\ \bibinfo {author}
  {\bibfnamefont {S.}~\bibnamefont {Kuhr}},\ }\href
  {http://dx.doi.org/10.1038/nphys3403} {\bibfield  {journal} {\bibinfo
  {journal} {Nat Phys}\ }\textbf {\bibinfo {volume} {11}},\ \bibinfo {pages}
  {738} (\bibinfo {year} {2015})},\ \bibinfo {note} {letter}\BibitemShut
  {NoStop}%
\bibitem [{\citenamefont {Edge}\ \emph {et~al.}(2015)\citenamefont {Edge},
  \citenamefont {Anderson}, \citenamefont {Jervis}, \citenamefont {McKay},
  \citenamefont {Day}, \citenamefont {Trotzky},\ and\ \citenamefont
  {Thywissen}}]{k3}%
  \BibitemOpen
  \bibfield  {author} {\bibinfo {author} {\bibfnamefont {G.~J.~A.}\
  \bibnamefont {Edge}}, \bibinfo {author} {\bibfnamefont {R.}~\bibnamefont
  {Anderson}}, \bibinfo {author} {\bibfnamefont {D.}~\bibnamefont {Jervis}},
  \bibinfo {author} {\bibfnamefont {D.~C.}\ \bibnamefont {McKay}}, \bibinfo
  {author} {\bibfnamefont {R.}~\bibnamefont {Day}}, \bibinfo {author}
  {\bibfnamefont {S.}~\bibnamefont {Trotzky}}, \ and\ \bibinfo {author}
  {\bibfnamefont {J.~H.}\ \bibnamefont {Thywissen}},\ }\href {\doibase
  10.1103/PhysRevA.92.063406} {\bibfield  {journal} {\bibinfo  {journal} {Phys.
  Rev. A}\ }\textbf {\bibinfo {volume} {92}},\ \bibinfo {pages} {063406}
  (\bibinfo {year} {2015})}\BibitemShut {NoStop}%
\bibitem [{\citenamefont {Greif}\ \emph {et~al.}(2016)\citenamefont {Greif},
  \citenamefont {Parsons}, \citenamefont {Mazurenko}, \citenamefont {Chiu},
  \citenamefont {Blatt}, \citenamefont {Huber}, \citenamefont {Ji},\ and\
  \citenamefont {Greiner}}]{fermi_insulator1}%
  \BibitemOpen
  \bibfield  {author} {\bibinfo {author} {\bibfnamefont {D.}~\bibnamefont
  {Greif}}, \bibinfo {author} {\bibfnamefont {M.~F.}\ \bibnamefont {Parsons}},
  \bibinfo {author} {\bibfnamefont {A.}~\bibnamefont {Mazurenko}}, \bibinfo
  {author} {\bibfnamefont {C.~S.}\ \bibnamefont {Chiu}}, \bibinfo {author}
  {\bibfnamefont {S.}~\bibnamefont {Blatt}}, \bibinfo {author} {\bibfnamefont
  {F.}~\bibnamefont {Huber}}, \bibinfo {author} {\bibfnamefont
  {G.}~\bibnamefont {Ji}}, \ and\ \bibinfo {author} {\bibfnamefont
  {M.}~\bibnamefont {Greiner}},\ }\href {\doibase 10.1126/science.aad9041}
  {\bibfield  {journal} {\bibinfo  {journal} {Science}\ }\textbf {\bibinfo
  {volume} {351}},\ \bibinfo {pages} {953} (\bibinfo {year}
  {2016})}\BibitemShut {NoStop}%
\bibitem [{\citenamefont {Cheuk}\ \emph {et~al.}(2016)\citenamefont {Cheuk},
  \citenamefont {Nichols}, \citenamefont {Lawrence}, \citenamefont {Okan},
  \citenamefont {Zhang},\ and\ \citenamefont {Zwierlein}}]{fermi_insulator2}%
  \BibitemOpen
  \bibfield  {author} {\bibinfo {author} {\bibfnamefont {L.~W.}\ \bibnamefont
  {Cheuk}}, \bibinfo {author} {\bibfnamefont {M.~A.}\ \bibnamefont {Nichols}},
  \bibinfo {author} {\bibfnamefont {K.~R.}\ \bibnamefont {Lawrence}}, \bibinfo
  {author} {\bibfnamefont {M.}~\bibnamefont {Okan}}, \bibinfo {author}
  {\bibfnamefont {H.}~\bibnamefont {Zhang}}, \ and\ \bibinfo {author}
  {\bibfnamefont {M.~W.}\ \bibnamefont {Zwierlein}},\ }\href {\doibase
  10.1103/PhysRevLett.116.235301} {\bibfield  {journal} {\bibinfo  {journal}
  {Phys. Rev. Lett.}\ }\textbf {\bibinfo {volume} {116}},\ \bibinfo {pages}
  {235301} (\bibinfo {year} {2016})}\BibitemShut {NoStop}%
\bibitem [{\citenamefont {Mazurenko}\ \emph {et~al.}(2017)\citenamefont
  {Mazurenko}, \citenamefont {Chiu}, \citenamefont {Ji}, \citenamefont
  {Parsons}, \citenamefont {Kan{\'a}sz-Nagy}, \citenamefont {Schmidt},
  \citenamefont {Grusdt}, \citenamefont {Demler}, \citenamefont {Greif},\ and\
  \citenamefont {Greiner}}]{antiferromagnetic}%
  \BibitemOpen
  \bibfield  {author} {\bibinfo {author} {\bibfnamefont {A.}~\bibnamefont
  {Mazurenko}}, \bibinfo {author} {\bibfnamefont {C.~S.}\ \bibnamefont {Chiu}},
  \bibinfo {author} {\bibfnamefont {G.}~\bibnamefont {Ji}}, \bibinfo {author}
  {\bibfnamefont {M.~F.}\ \bibnamefont {Parsons}}, \bibinfo {author}
  {\bibfnamefont {M.}~\bibnamefont {Kan{\'a}sz-Nagy}}, \bibinfo {author}
  {\bibfnamefont {R.}~\bibnamefont {Schmidt}}, \bibinfo {author} {\bibfnamefont
  {F.}~\bibnamefont {Grusdt}}, \bibinfo {author} {\bibfnamefont
  {E.}~\bibnamefont {Demler}}, \bibinfo {author} {\bibfnamefont
  {D.}~\bibnamefont {Greif}}, \ and\ \bibinfo {author} {\bibfnamefont
  {M.}~\bibnamefont {Greiner}},\ }\href@noop {} {\bibfield  {journal} {\bibinfo
   {journal} {Nature}\ }\textbf {\bibinfo {volume} {545}},\ \bibinfo {pages}
  {462} (\bibinfo {year} {2017})}\BibitemShut {NoStop}%
\bibitem [{\citenamefont {Rossini}\ and\ \citenamefont
  {Fazio}(2012)}]{extended_bhm1}%
  \BibitemOpen
  \bibfield  {author} {\bibinfo {author} {\bibfnamefont {D.}~\bibnamefont
  {Rossini}}\ and\ \bibinfo {author} {\bibfnamefont {R.}~\bibnamefont
  {Fazio}},\ }\href {http://stacks.iop.org/1367-2630/14/i=6/a=065012}
  {\bibfield  {journal} {\bibinfo  {journal} {New Journal of Physics}\ }\textbf
  {\bibinfo {volume} {14}},\ \bibinfo {pages} {065012} (\bibinfo {year}
  {2012})}\BibitemShut {NoStop}%
\bibitem [{\citenamefont {Baier}\ \emph {et~al.}(2016)\citenamefont {Baier},
  \citenamefont {Mark}, \citenamefont {Petter}, \citenamefont {Aikawa},
  \citenamefont {Chomaz}, \citenamefont {Cai}, \citenamefont {Baranov},
  \citenamefont {Zoller},\ and\ \citenamefont {Ferlaino}}]{extended_bhm2}%
  \BibitemOpen
  \bibfield  {author} {\bibinfo {author} {\bibfnamefont {S.}~\bibnamefont
  {Baier}}, \bibinfo {author} {\bibfnamefont {M.~J.}\ \bibnamefont {Mark}},
  \bibinfo {author} {\bibfnamefont {D.}~\bibnamefont {Petter}}, \bibinfo
  {author} {\bibfnamefont {K.}~\bibnamefont {Aikawa}}, \bibinfo {author}
  {\bibfnamefont {L.}~\bibnamefont {Chomaz}}, \bibinfo {author} {\bibfnamefont
  {Z.}~\bibnamefont {Cai}}, \bibinfo {author} {\bibfnamefont {M.}~\bibnamefont
  {Baranov}}, \bibinfo {author} {\bibfnamefont {P.}~\bibnamefont {Zoller}}, \
  and\ \bibinfo {author} {\bibfnamefont {F.}~\bibnamefont {Ferlaino}},\ }\href
  {\doibase 10.1126/science.aac9812} {\bibfield  {journal} {\bibinfo  {journal}
  {Science}\ }\textbf {\bibinfo {volume} {352}},\ \bibinfo {pages} {201}
  (\bibinfo {year} {2016})}\BibitemShut {NoStop}%
\bibitem [{\citenamefont {Yamamoto}\ \emph {et~al.}(2016)\citenamefont
  {Yamamoto}, \citenamefont {Kobayashi}, \citenamefont {Kuno}, \citenamefont
  {Kato},\ and\ \citenamefont {Takahashi}}]{ytterbium_narrow}%
  \BibitemOpen
  \bibfield  {author} {\bibinfo {author} {\bibfnamefont {R.}~\bibnamefont
  {Yamamoto}}, \bibinfo {author} {\bibfnamefont {J.}~\bibnamefont {Kobayashi}},
  \bibinfo {author} {\bibfnamefont {T.}~\bibnamefont {Kuno}}, \bibinfo {author}
  {\bibfnamefont {K.}~\bibnamefont {Kato}}, \ and\ \bibinfo {author}
  {\bibfnamefont {Y.}~\bibnamefont {Takahashi}},\ }\href
  {http://stacks.iop.org/1367-2630/18/i=2/a=023016} {\bibfield  {journal}
  {\bibinfo  {journal} {New Journal of Physics}\ }\textbf {\bibinfo {volume}
  {18}},\ \bibinfo {pages} {023016} (\bibinfo {year} {2016})}\BibitemShut
  {NoStop}%
\bibitem [{\citenamefont {Miranda}\ \emph {et~al.}(2015)\citenamefont
  {Miranda}, \citenamefont {Inoue}, \citenamefont {Okuyama}, \citenamefont
  {Nakamoto},\ and\ \citenamefont {Kozuma}}]{mm2}%
  \BibitemOpen
  \bibfield  {author} {\bibinfo {author} {\bibfnamefont {M.}~\bibnamefont
  {Miranda}}, \bibinfo {author} {\bibfnamefont {R.}~\bibnamefont {Inoue}},
  \bibinfo {author} {\bibfnamefont {Y.}~\bibnamefont {Okuyama}}, \bibinfo
  {author} {\bibfnamefont {A.}~\bibnamefont {Nakamoto}}, \ and\ \bibinfo
  {author} {\bibfnamefont {M.}~\bibnamefont {Kozuma}},\ }\href {\doibase
  10.1103/PhysRevA.91.063414} {\bibfield  {journal} {\bibinfo  {journal} {Phys.
  Rev. A}\ }\textbf {\bibinfo {volume} {91}},\ \bibinfo {pages} {063414}
  (\bibinfo {year} {2015})}\BibitemShut {NoStop}%
\bibitem [{\citenamefont {Miranda}\ \emph {et~al.}(2012)\citenamefont
  {Miranda}, \citenamefont {Nakamoto}, \citenamefont {Okuyama}, \citenamefont
  {Noguchi}, \citenamefont {Ueda},\ and\ \citenamefont {Kozuma}}]{mm1}%
  \BibitemOpen
  \bibfield  {author} {\bibinfo {author} {\bibfnamefont {M.}~\bibnamefont
  {Miranda}}, \bibinfo {author} {\bibfnamefont {A.}~\bibnamefont {Nakamoto}},
  \bibinfo {author} {\bibfnamefont {Y.}~\bibnamefont {Okuyama}}, \bibinfo
  {author} {\bibfnamefont {A.}~\bibnamefont {Noguchi}}, \bibinfo {author}
  {\bibfnamefont {M.}~\bibnamefont {Ueda}}, \ and\ \bibinfo {author}
  {\bibfnamefont {M.}~\bibnamefont {Kozuma}},\ }\href {\doibase
  10.1103/PhysRevA.86.063615} {\bibfield  {journal} {\bibinfo  {journal} {Phys.
  Rev. A}\ }\textbf {\bibinfo {volume} {86}},\ \bibinfo {pages} {063615}
  (\bibinfo {year} {2012})}\BibitemShut {NoStop}%
\bibitem [{\citenamefont {Greiner}\ \emph {et~al.}(2002)\citenamefont
  {Greiner}, \citenamefont {Mandel}, \citenamefont {Esslinger}, \citenamefont
  {Hansch},\ and\ \citenamefont {Bloch}}]{Greiner2002}%
  \BibitemOpen
  \bibfield  {author} {\bibinfo {author} {\bibfnamefont {M.}~\bibnamefont
  {Greiner}}, \bibinfo {author} {\bibfnamefont {O.}~\bibnamefont {Mandel}},
  \bibinfo {author} {\bibfnamefont {T.}~\bibnamefont {Esslinger}}, \bibinfo
  {author} {\bibfnamefont {T.~W.}\ \bibnamefont {Hansch}}, \ and\ \bibinfo
  {author} {\bibfnamefont {I.}~\bibnamefont {Bloch}},\ }\href {\doibase
  10.1038/415039a} {\bibfield  {journal} {\bibinfo  {journal} {Nature}\
  }\textbf {\bibinfo {volume} {415}},\ \bibinfo {pages} {39} (\bibinfo {year}
  {2002})}\BibitemShut {NoStop}%
\bibitem [{Note1()}]{Note1}%
  \BibitemOpen
  \bibinfo {note} {The superfluid-to-Mott-insulator transition is expected to
  occur at a lattice depth of $14\protect \tmspace +\thinmuskip {.1667em} E_r$.
  At the transition point, the tunneling rate is $J/h=7.7\protect \tmspace
  +\thinmuskip {.1667em}\protect \text {Hz}$, the interaction energy is
  $U/k_B=6.1\protect \tmspace +\thinmuskip {.1667em}\protect \text {nK}$ and
  the lattice transverse confinement is $\omega / 2\pi = 35\protect \tmspace
  +\thinmuskip {.1667em}\protect \text {Hz}$. At $26\protect \tmspace
  +\thinmuskip {.1667em} E_r$ the tunneling rate becomes negligible compared
  with the on-site interaction ($U/J \approx 240$) and fluctuations in the atom
  number are drastically reduced.}\BibitemShut {Stop}%
\bibitem [{\citenamefont {Tojo}\ \emph {et~al.}(2006)\citenamefont {Tojo},
  \citenamefont {Kitagawa}, \citenamefont {Enomoto}, \citenamefont {Kato},
  \citenamefont {Takasu}, \citenamefont {Kumakura},\ and\ \citenamefont
  {Takahashi}}]{556_PA}%
  \BibitemOpen
  \bibfield  {author} {\bibinfo {author} {\bibfnamefont {S.}~\bibnamefont
  {Tojo}}, \bibinfo {author} {\bibfnamefont {M.}~\bibnamefont {Kitagawa}},
  \bibinfo {author} {\bibfnamefont {K.}~\bibnamefont {Enomoto}}, \bibinfo
  {author} {\bibfnamefont {Y.}~\bibnamefont {Kato}}, \bibinfo {author}
  {\bibfnamefont {Y.}~\bibnamefont {Takasu}}, \bibinfo {author} {\bibfnamefont
  {M.}~\bibnamefont {Kumakura}}, \ and\ \bibinfo {author} {\bibfnamefont
  {Y.}~\bibnamefont {Takahashi}},\ }\href {\doibase
  10.1103/PhysRevLett.96.153201} {\bibfield  {journal} {\bibinfo  {journal}
  {Phys. Rev. Lett.}\ }\textbf {\bibinfo {volume} {96}},\ \bibinfo {pages}
  {153201} (\bibinfo {year} {2006})}\BibitemShut {NoStop}%
\bibitem [{\citenamefont {Sugawa}\ \emph {et~al.}(2011)\citenamefont {Sugawa},
  \citenamefont {Inaba}, \citenamefont {Taie}, \citenamefont {Yamazaki},
  \citenamefont {Yamashita},\ and\ \citenamefont {Takahashi}}]{556_PA2}%
  \BibitemOpen
  \bibfield  {author} {\bibinfo {author} {\bibfnamefont {S.}~\bibnamefont
  {Sugawa}}, \bibinfo {author} {\bibfnamefont {K.}~\bibnamefont {Inaba}},
  \bibinfo {author} {\bibfnamefont {S.}~\bibnamefont {Taie}}, \bibinfo {author}
  {\bibfnamefont {R.}~\bibnamefont {Yamazaki}}, \bibinfo {author}
  {\bibfnamefont {M.}~\bibnamefont {Yamashita}}, \ and\ \bibinfo {author}
  {\bibfnamefont {Y.}~\bibnamefont {Takahashi}},\ }\href {\doibase
  10.1038/nphys2028} {\bibfield  {journal} {\bibinfo  {journal} {Nat Phys}\
  }\textbf {\bibinfo {volume} {7}},\ \bibinfo {pages} {642} (\bibinfo {year}
  {2011})}\BibitemShut {NoStop}%
\bibitem [{\citenamefont {DeMarco}\ \emph {et~al.}(2005)\citenamefont
  {DeMarco}, \citenamefont {Lannert}, \citenamefont {Vishveshwara},\ and\
  \citenamefont {Wei}}]{shell1}%
  \BibitemOpen
  \bibfield  {author} {\bibinfo {author} {\bibfnamefont {B.}~\bibnamefont
  {DeMarco}}, \bibinfo {author} {\bibfnamefont {C.}~\bibnamefont {Lannert}},
  \bibinfo {author} {\bibfnamefont {S.}~\bibnamefont {Vishveshwara}}, \ and\
  \bibinfo {author} {\bibfnamefont {T.-C.}\ \bibnamefont {Wei}},\ }\href
  {\doibase 10.1103/PhysRevA.71.063601} {\bibfield  {journal} {\bibinfo
  {journal} {Phys. Rev. A}\ }\textbf {\bibinfo {volume} {71}},\ \bibinfo
  {pages} {063601} (\bibinfo {year} {2005})}\BibitemShut {NoStop}%
\bibitem [{\citenamefont {F\"olling}\ \emph {et~al.}(2006)\citenamefont
  {F\"olling}, \citenamefont {Widera}, \citenamefont {M\"uller}, \citenamefont
  {Gerbier},\ and\ \citenamefont {Bloch}}]{shell2}%
  \BibitemOpen
  \bibfield  {author} {\bibinfo {author} {\bibfnamefont {S.}~\bibnamefont
  {F\"olling}}, \bibinfo {author} {\bibfnamefont {A.}~\bibnamefont {Widera}},
  \bibinfo {author} {\bibfnamefont {T.}~\bibnamefont {M\"uller}}, \bibinfo
  {author} {\bibfnamefont {F.}~\bibnamefont {Gerbier}}, \ and\ \bibinfo
  {author} {\bibfnamefont {I.}~\bibnamefont {Bloch}},\ }\href {\doibase
  10.1103/PhysRevLett.97.060403} {\bibfield  {journal} {\bibinfo  {journal}
  {Phys. Rev. Lett.}\ }\textbf {\bibinfo {volume} {97}},\ \bibinfo {pages}
  {060403} (\bibinfo {year} {2006})}\BibitemShut {NoStop}%
\bibitem [{Note2()}]{Note2}%
  \BibitemOpen
  \bibinfo {note} {We estimate that the probability of finding two atoms
  occupying adjacent sites is $0.5\%$ for a square lattice comprised of
  $30\times 30$ sites.}\BibitemShut {Stop}%
\bibitem [{\citenamefont {Wessel}\ \emph {et~al.}(2004)\citenamefont {Wessel},
  \citenamefont {Alet}, \citenamefont {Troyer},\ and\ \citenamefont
  {Batrouni}}]{bhm1}%
  \BibitemOpen
  \bibfield  {author} {\bibinfo {author} {\bibfnamefont {S.}~\bibnamefont
  {Wessel}}, \bibinfo {author} {\bibfnamefont {F.}~\bibnamefont {Alet}},
  \bibinfo {author} {\bibfnamefont {M.}~\bibnamefont {Troyer}}, \ and\ \bibinfo
  {author} {\bibfnamefont {G.~G.}\ \bibnamefont {Batrouni}},\ }\href {\doibase
  10.1103/PhysRevA.70.053615} {\bibfield  {journal} {\bibinfo  {journal} {Phys.
  Rev. A}\ }\textbf {\bibinfo {volume} {70}},\ \bibinfo {pages} {053615}
  (\bibinfo {year} {2004})}\BibitemShut {NoStop}%
\bibitem [{\citenamefont {Hofrichter}\ \emph {et~al.}(2016)\citenamefont
  {Hofrichter}, \citenamefont {Riegger}, \citenamefont {Scazza}, \citenamefont
  {H\"ofer}, \citenamefont {Fernandes}, \citenamefont {Bloch},\ and\
  \citenamefont {F\"olling}}]{fermi_sun1}%
  \BibitemOpen
  \bibfield  {author} {\bibinfo {author} {\bibfnamefont {C.}~\bibnamefont
  {Hofrichter}}, \bibinfo {author} {\bibfnamefont {L.}~\bibnamefont {Riegger}},
  \bibinfo {author} {\bibfnamefont {F.}~\bibnamefont {Scazza}}, \bibinfo
  {author} {\bibfnamefont {M.}~\bibnamefont {H\"ofer}}, \bibinfo {author}
  {\bibfnamefont {D.~R.}\ \bibnamefont {Fernandes}}, \bibinfo {author}
  {\bibfnamefont {I.}~\bibnamefont {Bloch}}, \ and\ \bibinfo {author}
  {\bibfnamefont {S.}~\bibnamefont {F\"olling}},\ }\href {\doibase
  10.1103/PhysRevX.6.021030} {\bibfield  {journal} {\bibinfo  {journal} {Phys.
  Rev. X}\ }\textbf {\bibinfo {volume} {6}},\ \bibinfo {pages} {021030}
  (\bibinfo {year} {2016})}\BibitemShut {NoStop}%
\bibitem [{\citenamefont {Honerkamp}\ and\ \citenamefont
  {Hofstetter}(2004)}]{fermi_sun2}%
  \BibitemOpen
  \bibfield  {author} {\bibinfo {author} {\bibfnamefont {C.}~\bibnamefont
  {Honerkamp}}\ and\ \bibinfo {author} {\bibfnamefont {W.}~\bibnamefont
  {Hofstetter}},\ }\href {\doibase 10.1103/PhysRevLett.92.170403} {\bibfield
  {journal} {\bibinfo  {journal} {Phys. Rev. Lett.}\ }\textbf {\bibinfo
  {volume} {92}},\ \bibinfo {pages} {170403} (\bibinfo {year}
  {2004})}\BibitemShut {NoStop}%
\end{thebibliography}%

\end{document}